# POSTULATE-BASED PROOF OF THE P ≠ NP HYPOTHESIS

O.V. German

The Belarusian state university of Informatics and Radioelectronics
220600, P. Brovki,6
Minsk, Republic of Belarus
e-mail: ovgerman@tut.by

**Abstract**

The paper contains a proof for the P ≠ NP hypothesis with the help of the two «natural» postulates. The postulates restrict capacity of the Turing machines and state that each independent and necessary condition of the problem should be considered by a solver (Turing machine) individually, not in groups. That is, a solver should spend at least one step to deal with the condition and, therefore, if the amount of independent conditions is exponentially growing with polynomially growing problem sizes then exponential time is needed to find a solution. With the postulates, it is enough to build a natural (not pure mathematical) proof that P ≠ NP.

## 1 INTRODUCTION

In this paper, we show that under quite a general supposition no efficient algorithm exists for SATISFIABILITY problem [1] (SAT, for short). The supposition is based on two hypotheses. The first one is the next: when working with problem conditions the solver (Turing machine) «takes into account» each independent and necessary condition separately from the others (that is, individually, spending at least one step for it). It is beyond our interests the solutions previously found for the problem(s). So, we are interested only in algorithms which find (initially unknown) solutions from the problem specification represented by some set of formulas. The solution process should warrant correctness of the result for each individual problem of the given type.

The second hypothesis sounds like this: no Turing machine (TM) exists with throughput exceeding some constant value (this is similar to the well-known restriction on the light speed in relativity theory).
These hypotheses are sufficient to prove the inequality between P and NP.

To prove P≠NP, a number of significant efforts were undertaken. A short review of them may be found in [2]. Thus, the methods of diagonalization and relativization were applied by analogy with the same methods used to prove the undecidability of some well-known algorithmic problems (HALT, for example). However, as stated in [3], there are different relativizations addmitting both P = NP and P ≠ NP. One should note a schematic approach by A. Razborov [4, 5] who showed superpolynomial complexity of functional schemes in the AND, OR basis to realize a solver for CLIQUE problem. There are no encouring results in the AND, OR, NOT basis as well. The best complexity estimation for a 3-SAT problem is $1.5^n$ [6] where $n$ is the numver of variables. Other complexity estimations and approaches may be found in [7].

There exists a viewpoint that P ≠?NP problem is not a pure mathematical and, in particular, does not depend on the axioms of Zermelo-Fraenkel set theory. In [8], the discrepancy between polynomial-time language recognition algorithm and total polynomiality of the recognizer is showed. The result may points to unresolveability of the P?=NP problem within frameworks of the standard mathematics. Let us reproduce here the main idea of [8]. Consider language PLAR consisting of the words $y = y_1 y_2$, where

1. $y_1$ denotes a specification (rule set) of arbitrary Turing machine $MT_\eta$.
2. $y_2$ represents a proof of the statement asserting that $MT_\eta$, defined by the word $y_1$, declines any word $y$, starting with $y_1$ (that is, $y = y_1 Z$ with arbitrary $Z$), for a time restricted by some fixed polynomial of the length of the word $y = y_1 Z$.

Clearly, PLAR is not empty as it contains at least one word for a Turing machine which declines any input for one step. Let us build Turing machine $MT_{PLAR}$ to recognize PLAR for polynomial time. This machine should test the proof $y_2$. This requires polynomial time. Consequently, if $y_2$ is correct then $MT_{PLAR}$ accepts word $y = y_1 @ y_2$. One can prove the next

**Theorem A**. $MT_{PLAR}$ recognizes PLAR for polynomial time.

**Theorem B**. It is impossible to prove that $MT_{PLAR}$ is a totally polinomial-time machine.

Let there be some proof $D_{polLAR}$ of total polynomiality of $MT_{PLAR}$. Then $MT_{PLAR}$ declines any word of the type $y_1^{LAR}Z$ for polynomial time. This gives a proof $\vartheta$ as a direct consequence from $D_{polLAR}$. Hence, $MT_{PLAR}$ should accept word $y_1^{LAR}\vartheta$ what leads to a contradiction.

The problem in this proof lays in the fact that all proofs represented by the words $y_2$ form some fixed infinite set, while the proof for Theorem B is added to this complete set. This means, that any «proof» of total polynomiality of $MT_{PLAR}$ is an inference, not a standard mathematical proof.

A lot of papers were published with the proofs both of P = NP and P≠NP [9-11] (see also web-site [12]). Many proofs, especially for the statement P = NP were afterwards recognized as incorrect. Some proofs were left without attention. Nevertheless, the P =?NP problem cannot be arbitrarily closed from the further attempts to solve it.

One could seek a solution with the help of phisycal concepts. The idea to unite mathematical and physical approaches is presented, for example, in [13]. It is necessary to bear in mind that a Turing machine is not only an abstract mathematical notion, but also an information processing device. Due to this, physical analogs get obviously important sense. This relates, for instance, to the notions of information quantity, entropy, throughput of the computational device, and some others. These notions have, first of all, a physical sense. Namely from these positions with application of necessary mathematical means, the given paper is built. The postulates accepted here, have both mathematical and physical nature, that is why we do not claim that our approach contains a pure mathematical proof of the P≠ NP problem. But mathematical substantiation of the approach is a part of the common approach outlined in the pioneer paper [14] which we will follow below.

## 2. FORMALIZATIONS

**Definition.** A condition is a formula (formal expression) with variables. A condition can be true or false with respect to its variable values.

**Definition.** A problem is a non-empty set of conditions. (An arbitrary) solution to the problem is a set of values of its variables satisfying all the explicitly stated conditions of the problem.

When solving a problem, a solver (Turing machine) works with the problem conditions. A part of conditions is stated explicitly, while the other part of conditions is hidden in the problem formulation. Each condition links the solution elements.

A condition is taken into consideration, provided the solver makes one or more steps to test if it is true or false, or treats it in the solution pocess.

A condition that does not logically follows from the other conditions is called independent of those ones.

If the falsehood of a condition leads to the loss of all solutions to the problem, then this condition is called necessary.

**Definition.** Any necessary and sufficient set of conditions of a problem $A$, which should be taken into consideration by its solver, is called infological set of this problem and designated by $\text{Inf}_A^{\text{set}}$. The elements of $\text{Inf}_A^{\text{set}}$ are called the infs. In general, the problem may have more than one infological set.

If a truth value (true or false) of some condition does not influence on the correctness of a problem solution, then this condition is not included at least in one infological set of the problem. If non-fulfillment of a condition leads to loss of all problem solutions, then this condition is included into each infological set or can be inferred from it. Provided, that a condition is independent of the other ones, it should be explicitly presented in infological set as we require of infs.

In what follows, we shall deal only with those conditions (infs) which are necessary to find a solution and independent.

We are interested only in the algorithms, which take infs into consideration, that is, not ignore them, and shall build a problem not permitting ignoring the infs.

In this paper, we deal with the problems formulated with $n > 1$ Boolean variables $x_1, x_2, \ldots, x_n$ (or integer-valued variables taking values from the restricted diapazons). A solution to the problem is represented by some feasible (satisfying) interpretation (a set of $x_1, x_2, \ldots, x_n$ values meeting the problem conditions stated in its specification). This somewhat differs such type of problems from YES-NO problems, however, not essentially. A principal point consists in the following: whether it is sufficient to find any one feasible solution (satisfying interpretation), if it exists, or it is required to find all feasible solutions.

**Definition.** A *generating scheme* represents a rule (rules), pointing to how to compute each solution to the problem.

**Definition.** The problems requiring to find any one feasible (valid) solution (if exists) will be called *ASP* (*a*ny *s*olution *p*roblems), while the problems requiring to find some generating scheme to find all solutions, will be called *TSP* (*t*otal *s*olution *p*roblems).

It is clear that if *TSP* has exponentially growing set of feasible solutions then even their simple enumeration requires an exponential time expenses. However, the answer to *TSP* may be in some cases represented as a set of generating rules (substitutions) producing values for the problem variables. To be clear, let us consider a problem with integer-valued variables $y_i$:

$$52y_1 - 15y_2 - 3y_3 = 2,$$
$$y_1 \in [-2, 2],$$
$$y_2 \in [-3, 7],$$
$$y_3 \in [-23, 2], \text{ all } y_i \text{ are integer.}$$

In the example, a required substitution system has the next possible form:
$$x_1 + x_2 + x_3 = 2,$$
$$y_1 = x_1 + x_2 - 2x_3,$$
$$y_2 = 3x_1 + 4x_2 - 3x_3,$$
$$y_3 = 2x_1 - 3x_2 - 20x_3,$$

$$x_i \in [0,1], \text{ all } x_i \text{ are integer.}$$

Additionally, it is required that:

**R1)** the smallest (the largest) value of $L_i$ ($R_i$) for each variable $y_i$ obviously is not less (greater) than the sum of all negative (positive) coefficients in the corresponding substitution for $y_i$ or 0, if there are no such coefficients.

**R2)** substitutions for $x_i$ are being looked for in a strict order from the substitution system accordingly to Gauss direct method of exclusions [15, vol.1, p.53] as follows: $x_1$ is expressed from the substitution for $y_1$, and $x_1$ should have coefficient $+1$ or $-1$ in this expression. Then expression for $x_1$ is used instead of $x_1$ in the expressions for $y_2$, ..., $y_n$. Then one takes expression for $y_2$ and gets substitution for $x_2$ from it. Again, $x_2$ should have coefficient $+1$ or $-1$. The obtained substitution for $x_2$ is then used in expressions for $y_3$, ..., $y_n$. The process repeats by analogy for the rest variables $x_i$, $i = 3, 4, ..., n$ by replacing $x_i$ in $y_{i+1}$, ..., $y_n$. Each time $x_i$ should have coefficient either $+1$ or $-1$ in the expression for $y_i$. The obtained substitutions should contain integer coefficients only. Thus, in the example we have

$$x_1 = y_1 - x_2 + 2x_3; \quad x_2 = -3y_1 + y_2 - 3x_3; \quad x_3 = 17y_1 - 5y_2 - y_3.$$

Clearly, from the obtained substitutions with the help of the backward Gauss method one can find the integer-valued substitutions for each $x_i$ with variables $y_j$ ($i, j = 1, ..., n$) only:

$$x_1 = 89y_1 - 26y_2 - 5y_3; \, x_2 = -54y_1 + 16y_2 + 3y_3; \, x_3 = 17y_1 - 5y_2 - y_3.$$

Because of the crucial importance of this problem, let us denote $A = \{a_1, a_2, ..., a_n\}$, $\boldsymbol{y} = \{y_1, y_2, ..., y_n\}$, $\boldsymbol{x} = \{x_1, x_2, ..., x_n\}$, $\boldsymbol{d} = \{d_i | d_i = [L_i, R_i]\}$, and use the abbreviation SUBST($\boldsymbol{y}$, $\boldsymbol{x}$, $A$, $\boldsymbol{d}$, $n$, $c$, $\Omega$) to denote its specification with the following formulation. Let there be given

$$\begin{aligned} a_1 y_1 + a_2 y_2 + ... + a_n y_n = c, \\ y_i \in [L_i, R_i], \quad y_i, a_i, L_i, R_i, c \text{ - all integer.} \end{aligned} \quad (1)$$

with known $a_i, L_i, R_i, c$ and unknown $y_i$, $i = 1, ..., n$. Denote the length of the specification (**1**) by $LG_3$. Then it is asked, if for the given (fixed) polynomial $\Omega$ there exist efficiently verifiable condition and the system of substitutions of the type

$$\begin{aligned} & x_1 + x_2 + ... + x_n = c \\ & x_i \in [0,1], \\ & y_i = b_{i1} x_1 + b_{i2} x_2 + ... + b_{in} x_n, \quad i = 1,...,n, \\ & \text{all } b_{ij} \text{ are integer,} \end{aligned} \quad (2)$$

and, besides,

1) the sizes of the substitution matrix $B = [b_{i,j}]$ and the sizes of the inverse matrix $B^{-1} = [b^*_{i,j}]$ do not exceed $\Omega(LG_3)$;

2) **R1**, **R2** are satisfied.

**Definition** [15]. The sizes of a rational number $\gamma = p/q$ ($p$, $q$ – are integer and coprime numbers, $q \neq 0$), rational vector $c = (\gamma_1, \gamma_2, ..., \gamma_n)$ and rational matrix $B = [b_{ij}]$ are defined as follows

$$\mathbf{size}(\gamma) = 1 + \lceil \log_2(|p|+1) \rceil + \lceil \log_2(|q|+1) \rceil,$$

$$\mathbf{size}(c) = n + \mathbf{size}(\gamma_1) + ... + \mathbf{size}(\gamma_n),$$

$$\mathbf{size}(B) = m \cdot n + \Sigma_{i,j} \, \mathbf{size}(b_{ij}),$$

where $\lceil x \rceil$ – is minimal integer value greater than $x$.

**Theorem 1**. If for conditions (**1**) there exists system (**2**), then for each integer-valued set, satisfying (**1**), there exists a unique integer-valued set, satisfying (**2**), and vice versa.

**Proof**. Can be simply obtained from linear algebra provided that matrix $B = [b_{i,j}]$ is not singular. The requirement of integrality of the coefficients is fulfilled by **R1**, **R2**.

**Note 1**. The number of satisfiable problems SUBST($y$, $x$, $A$, $d$, $n$, $c$, $\Omega$) is infinite. Elementary technique to generate them consists in the following. Take each substitution $y_i = b_{i1}x_1 + b_{i2}x_2 + ... + b_{in}x_n$ starting with $i = 1$ and set $b_{11} = 1$ with the other coefficients $b_{1j}$ ($j > 1$) representing arbitrary integer values. For $i = 2$, the coefficients $b_{2j}$ ($j <> 2$) represents arbitrary integer values, besides $b_{22}$. The value of $b_{22}$ is defined in such a way that after performing substitution for $x_1$ from $y_1 = x_1 + b_{12}x_2 + ... + b_{1n}x_n$ to make $b_{22} = 1$ or $b_{22} = -1$ and so on.

**Note 2**. Requirement **R2** is not peculiar. One can restrict the substitutions for $x_i$ by $y_j$ ($i, j = 1, ..., n$) only by integer-valued ones, since it is possible to show how an arbitrary integer-valued substitution matrix can be reduced to the form meeting **R2**. For this aim, one should apply the known technique on the basis of Euclidean method for seeking the integer-valued solutions of the linear algebraic equalities with multiple variables and integer (rational) coefficients (see, for instance [16, p.p. 52–53]). However, the technique, outlined above, is further used to estimate the memory expences for representation of the coefficients of the substitution matrix $B$ and its inverse matrix $B^{-1}$.

It is clear, that from system (**2**) the values of $x_i$ are defined elementary and deliver the corresponding solutions to the original problem.

**Theorem 2.** A SUBST($y$, $x$, $A$, $d$, $n$, $c$, $\Omega$) is polynomially reducible to SATISFIABILITY problem.

**Proof**. Can be found in [14].

It is necessary to note that by means of the given polynomial $\Omega$, one is in position to efficiently define the sizes of the Boolean representation of the integer-valued coefficients of substitution matrix $B$ and its inverse matrix $B^{-1}$. There remains to point to the fact that maximal sizes of the intermediate coefficients obtained by procedure for verifying the requiremeent **R2**, are restricted by the value 4·**size**($B$) [15, vol. 1, p.56] where **size**($B$) defines the sizes of the substitution matrix $B$.

The values of intermediate coefficients can be found from the relationships pointed to in Gantmaher's book [17, p.43] and are expressed through the minors of the matrix $B$. If $M$ is a maximal absolute value of a coefficient in substitution matrix $B$, then the value of each minor in $B$ is not higher than $n! \cdot M^n$ what requires no more than $n \cdot Log_2(n \cdot M)$ bits for representation in memory, i.e. is estimated as $O(\mathbf{size}(B))$ with **size**($B$) not exceeding $n^2 \cdot \Omega(LG_3)$.

One can conclude that the technique used to reduce SUBST($y$, $x$, $A$, $d$, $n$, $c$, $\Omega$) to SAT is polynomially efficient. It follows then that if there not exist a polynomially efficient algorithm for SUBST($y$, $x$, $A$, $d$, $n$, $c$, $\Omega$) with some fixed $\Omega$ then SAT has no polynomial solution as well.

Next, we formulate two Postulates, which play decisive role for our goal.

# 3. THE POSTULATES

## POSTULATE 1.

Any *TSP*-problem requires to individually treat (take into account) each independent and necessary condition, spending for this at least one step of the solver (Turing machine − TM) work. That is, if the number of all independent and at the same time necessary conditions is $q$ then the number of steps, TM should perform, is not less than $q$.

A natural explanation of POSTULATE 1 may be given through the system of logical clauses (disjuncts) $D_1$, $D_2$, ..., $D_p$ forming a SAT problem. In fact, if some clause depends on the others (that is, logically follows from them) then that clause can be deleted (not taken into account) from SAT without loss of every solution what is necessary for *TSP*-problems. On the contrary, if a clause does not depend on the others than it cannot be deleted and should be taken into consideration individually or in groups. However, POSTULATE 1 does not admit the last opportunity. Indeed, let $r = p \mathbin{\&} q \to p$, with $p$ and $q$ mutually independent ($r$ is a group, consisting of the conditions $p$ and $q$). From the logical value (true/false) of $q$, one cannot establish logical value of $p$ without analyzing $p$ individually. By this, the value of the whole group $r$ cannot be established without knowing logical values of the conditions $p$ and $q$ from $r$.

Independence has a fundamental nature. If some condition $C$ is necessary and does not depend on the other conditions, then truth or falsity of those last says nothing about truth or falsity of $C$. Hence, $C$ should be taken into account by necessity. In a standard way, it is adopted that formula $C$ does not depend on the formulas $\varphi_1$, $\varphi_2$, ..., $\varphi_k$, provided that there exists some interpretation $I$, such that $\varphi_1(I) \mathbin{\&} \varphi_2(I) \mathbin{\&} ... \mathbin{\&} \varphi_k(I) = true$, but $C(I) = false$.

This definition of independence should be somewhat modified for the paper needs. Let $I_1$, $I_2$, ... , $I_n$ be particular interpretations and $I = I_1 I_2 ... I_n$ stand for their concatenation. Let $\varphi_1(I) = f(I_1)$, $\varphi_2(I) = f(I_2)$, ..., $\varphi_{n-1}(I) = f(I_{n-1})$, $\varphi_n(I) = f(I_n)$, and there exists interpretation $I = I_1 I_2 ... I_n$, in which $\varphi_1(I)$, $\varphi_2(I)$,..., $\varphi_{n-1}(I)$ all are true, and $\varphi_n(I)$ is false. Hence, $\varphi_n$ does not depend on $\varphi_1(I)$, $\varphi_2(I)$, ... , $\varphi_{n-1}(I)$.

This formulation of independence will be further referred to as independence in private interpretations. Evidently, independence in private interpretations is a particular case of the independence defined in standard way.

Finally, a condition $C$ is a necessary one, provided, that its failure (falsity) leads to the loss of all solutions.

Note that an independent condition is not obligatory a necessary one. A necessary condition may be dependent. The conditions explicitly formulated in a problem specification are necessary ones.

We shall use the next

**Lemma♣**. Let $\varphi \& F$ be a compatible system of logical formulas. Then from $\varphi \& F \rightarrow H$ follows $\varphi \& \neg H \rightarrow \neg F$ ($\neg$ denotes logical negation).

Proof folows from equivalence $x \rightarrow y \equiv \neg x \vee y$.

Definition of reduction of a problem $A$ to problem $B$ can be found in [1].

Denote by $TH$ the throughput of TM; by $N$ – the number of steps it performs in order to reach the final state; by $I$ – the number of the infs, TM takes into consideration during its work. Let $TH = I / N$.

## POSTULATE 2.

For each totally finite $TM_A$, and each common problem solved by $TM_A$, the following relationship is true

$$TH_A \leq c_A < \infty, \qquad (3)$$

where $c_A$ stands for some fixed constant value related to this $TM_A$.
From (**3**) one concludes that each physical data processing device cannot process an infinitely many infs (bits of information) per one step (cycle/transmission) of its work. This fundamental principle is asserted, in particular, in [18].

Let us briefly dwell on a possible counter-argument through the well-known acceleration theorem by M. Blum [19]. This theorem asserts that there exist general recursive functions $f$ with values from $\{0,1\}$ and such that for each TM $Z_i$, calculating $f(n)$ for $\Phi_i(n)$ steps, there exists another TM $Z_j$, which calculates $f(n)$ significantly faster for $\Phi_j(n)$ steps with $\Phi_i(n) > 2^{\Phi_j(n)}$.

Moreover, there is an infinite sequence $\tau$ of TMs, calculating *f*, in which for every neighbor pair of Turing machines $Z_s$ и $Z_{s+1}$ one has inequality $\Phi_s(n) > 2^{\Phi_{s+1}(n)}$, correct for almost all *n*.

From Blum's theorem, however, it does not follow existence of TM with unlimited throughput. Even if one accepts that TMs, computing the same function, process the equal quantity of information, the Blum's theorem only states for each pair $Z_s$ и $Z_{s+1}$ availability of the fixed number *m* defined for this pair of TMs, such that for almost each $n > m$ one has $\Phi_s(n) > 2^{\Phi_{s+1}(n)}$. This means that starting from the value *m*, time expences of $Z_s$ grows significantly faster than time expences of $Z_{s+1}$. The Blum's acceleration theorem does not impose restrictions on the upper boundary of $\Phi_i(n)$ for all indices *i* from $\tau$. Hence, for each fixed *n*, the fastest computation of *f(n)*, say on $Z_r$, may require an arbitrary many number of steps, while for $l > n$ the fastest computation of *f(l)* is performed by another TM, say, $Z_y$ with very great value of $\Phi_y(l)$ as well.

We have reached the point in our reasoning where it is required to introduce some consistent *TSP* problem A with exponentially growing sizes of $\text{Inf}_A^{\text{set}}$ for linearly grow of the variables number *n*. It requires of us to show that minimum Conjunctive Normal Form (CNF) of this problem grows exponentially in sizes provided, that the number of variables grows linearly. Consequently, this problem cannot be solved for polynomial time whichever solver is used, provided, that all problem infs are taken into consideration and each inf is considered individually, not in groups. The reader evidently has guessed that we intend to use SUBST(*y*, *x*, A, *d*, *n*, *c*, $\Omega$).

## 4. INFs

Some NP-complete problems use the condition formally represented as

$$a_1x_1 + a_2x_2 + ... + a_nx_n = c, \tag{4}$$

where $a_i$, *c* are integer non-negative numbers and $x_i \in \{0, 1\}$. In particular, some private case of (4) is treated in a Minimum-Size Covering Problem (MSCP) of a 0,1-matrix. SAT can be reduced to (**4**). Let us call (**4**) a Container Packing Problem (CPP). One can see that CPP is an NP-complete problem. Now ask, if CPP can be polynomially reduced to equivalent SAT provided, that both problems are specified with the same set of variables? The answer is delivered by

**Theorem 3.** It is impossible to reduce CPP($x_1$, $x_2$, …, $x_n$) to equivalent SAT($x_1$, $x_2$, …, $x_n$) with the sizes of SAT($x_1$, $x_2$, …, $x_n$) restricted by some polynomial of $n$.

**Proof**. See [14]. The proof uses the J.Robinson resolution technique [20] and is considered in detail in the referred paper.

## 5. THE SUBST($y$, $x$, $A$, $d$, $n$, $c$, $\Omega$) PROBLEM

Consider a system of substitutions

$$y_i = a_i + b_{i1}x_1 + b_{i2}x_2 + \ldots + b_{in}x_n, \quad i = 1, n, \tag{5}$$

with integer $a_i$, $b_{ij}$, and binary $x_i$. The main requirement to the substitutions (**5**) is to provide the uniqueness of the transformation, or in the other words, to ensure that each set $\boldsymbol{x}^* = <x^*_1, x^*_2, \ldots, x^*_n>$ is mapped to unique set $\boldsymbol{y}^* = <y^*_1, y^*_2, \ldots, y^*_n>$ (the opposite is obvious). The said requirement is fulfilled by nondegeneracy of the substitution matrix $B = [b_{ij}]$.

From (**5**),

$$\boldsymbol{y} = B^{-1} \cdot (\boldsymbol{x} - \boldsymbol{a}). \tag{6}$$

**Theorem 4** [15].

1. If the system of rational equations (**5**) is consistent then it has a solution $\boldsymbol{y}$ with the sizes restricted by some polynomial $\wp_1$ of the sizes of $B^{-1}$ and ($\boldsymbol{x} - \boldsymbol{a}$).

2. The inverse matrix $B^{-1}$ has the sizes, restricted by some polynomial $\wp_2$ of the matrix $B$ sizes.

Notice that we are interested in the values of $y_i$ which define the values of $x_i \in \{0,1\}$. For convinience, let us set $\boldsymbol{b} = 0$. From this, the sizes of $\boldsymbol{y}$ are restricted by some polynomial $\wp_3(\text{size}(B)) = \wp_1(\wp_2(\text{size}(B)))$.

One can see that for $m = n$ and some fixed constant $k$, **size**$(B) \leq O(k \cdot n^2 \cdot (\max_{i,j}$ **size**$(b_{ij})))$. So, there remains to provide that the value of **size**$(b_{ij})$ grows not faster than some fixed polynomial. However, this requirement is trivial as in selecting the coefficients of the matrix $B$ one should preserve only its nodegeneracy (det $B \neq 0$).

We have reached the final point. Accordingly to SUBST($y$, $x$, $A$, $d$, $n$, $c$, $\Omega$)**,** it is necessary to build a function $f$ (generated by the system of substitutions) mapping each satisfying interpretation $I(x^*)$ ($x$-values) satisfying system (**2**), to the unique interpretation $I(y^*)$ ($y$-values) satisfying system (**1**) or vice versa. It was demonstrated that the number of feasible interpretations for the system (**2**) grows exponentially with linear growth of the number of variables $n$. The inverse function $f^{-1}$ can be found (in the form of the inverse matrix $B^{-1}$) provided, that the matrix $B$ of substitutions is nondegenerate. Therefore, if to consider each pair of interpretations $(x^*, y^*)$ satisfying to (**1, 2**), separately from the other pairs $(x, y)$, then it is necessary to consider exponentially growing number of all pairs $(x, y)$, and by POSTULATE 2 to spend exponentially growing time to solve SUBST($y$, $x$, $A$, $d$, $n$, $c$, $\Omega$) in general. However, let us try to refute this conclusion and suppose that after establishing some part of all pairs $(x, y)$, satisfying (**1, 2**), the remaining pairs of interpretations may be not considered and, therefore, be ignored by the solver. Again, to make our reasoning clearer, consider an illustration. Let the generating rule be as before

$$x_1 + x_2 + x_3 + x_4 + x_5 = 3. \qquad (7)$$

The next table shows all feasible pairs of interpretatios, satisfying the systems (**1, 2**).

Table 1. The pairs of iterpretations deliverig the solutions to the systems (**1, 2**) (private case)

| $x_1$ | $x_2$ | $x_3$ | $x_4$ | $x_5$ | $y_1$ | $y_2$ | $y_3$ | $y_4$ | $y_5$ |
|---|---|---|---|---|---|---|---|---|---|
| 0 | 0 | 1 | 1 | 1 | $y_{11}$ | $y_{21}$ | $y_{31}$ | $y_{41}$ | $y_{51}$ |
| 0 | 1 | 0 | 1 | 1 | $y_{12}$ | $y_{22}$ | $y_{32}$ | $y_{42}$ | $y_{52}$ |
| 0 | 1 | 1 | 0 | 1 | $y_{13}$ | $y_{23}$ | $y_{33}$ | $y_{43}$ | $y_{53}$ |
| 0 | 1 | 1 | 1 | 0 | $y_{14}$ | $y_{24}$ | $y_{34}$ | $y_{44}$ | $y_{54}$ |
| 1 | 0 | 0 | 1 | 1 | $y_{15}$ | $y_{25}$ | $y_{35}$ | $y_{45}$ | $y_{55}$ |
| ... | ... | ... | ... | ... | | ... | ... | ... | ... |
| 1 | 1 | 1 | 0 | 0 | $y_{1,10}$ | $y_{2,10}$ | $y_{3,10}$ | $y_{4,10}$ | $y_{5,10}$ |

We assume that the rows of the table 1 are arranged in descending strong order of values in the column $y_1$. This assumption does not violate the strength of the results obtained. With the help of the rule **R2**, it is possible to generate an infinite number of the individual problems SUBST($y$, $x$, $A$, $d$, $n$, $c$, $\Omega$) satisfying this assumption. For this, one should use the first substitution

$$y_1 = b_{11}x_1 + b_{12}x_2 + ... + b_{1n}x_n \quad (8)$$

for $y_1$ with the coefficient $b_{11} = 1$ and each subsequent coefficient $b_{1j} > \sum_{k<j} b_{1k}$. It may be proved by induction on the number of variables in the substitution (**14**) that in this case the values of $y_1$ would be arranged in desceinding order (leave this to the reader). Consider, for example, the fifth row in the table 1

| 1 | 0 | 0 | 1 | 1 | $y_{15}$ | $y_{25}$ | $y_{35}$ | $y_{45}$ | $y_{55}$ |

Suppose that it depends on the previous rows in the table 1. This means, that the values $y_{15}$, $y_{25}$, $y_{35}$, $y_{45}$, $y_{55}$ automatically fall into ranges $[L_i, R_i]$ provided, that the previous rows <$y_{1i}$, $y_{2i}$, $y_{3i}$, $y_{4i}$, $y_{5i}$> have fallen into ranges $[L_i, R_i]$. However, before solving any individual problem SUBST($y$, $x$, $A$, $d$, $n$, $c$, $\Omega$) (including the considered one) this fact is unknown, and there exist two types of SUBST($y$, $x$, $A$, $d$, $n$, $c$, $\Omega$) such that $y_{15}$ falls into $[L_1, R_1]$ for the first type, and does not fall into $[L_1, R_1]$ for the second type of SUBST($y$, $x$, $A$, $d$, $n$, $c$, $\Omega$). Indeed, in comparison with the previous row <$y_{14}$, $y_{24}$, $y_{34}$, $y_{44}$, $y_{54}$> the following relations takes place $y_{15} < y_{14}$, $y_{14} \in [L_1, R_1]$. Then, by increasing the value of $L_1$ one can provide that $y_{11}$, $y_{12}$, $y_{13}$, $y_{14} \in [L_1, R_1]$, but $y_{15} \notin [L_1, R_1]$. This means that the row <$y_{15}$, $y_{25}$, $y_{35}$, $y_{45}$, $y_{55}$> should be taken into consideration by the solver even in the case that the previous rows satisfied the corresponding ranges $[L_i, R_i]$. Note that lexicographical descending order of the rows with interpretations can be obtained by a simple indeces mixing in variables $x_1$, $x_2$, $x_3$, $x_4$, $x_5$ as they are mutually independent.

The situation, we have described, is applicable to any part of the table with the interpretations (including the entire table) and means that each row of the table should be considered separately by virtue of POSTULATE 1 as a formula independent in particular interpretations from the other ones. Indeed, one can define the next formulas

$\varphi_1(I) = h(\text{SUBST}(y, x, A, d, n, c, \Omega), I_1)$, $\varphi_2(I) = h(\text{SUBST}(y, x, A, d, n, c, \Omega), I_2)$,..., $\varphi_n(I) = h(\text{SUBST}(y, x, A, d, n, c, \Omega)), I_n)$ where $h(\text{SUBST}(y, x, A, d, n, c, \Omega)), I_i)$ is true for the $i$-th row of the interpretation table 1 with $I_i = (x_i, y_i)$ ($x_i, y_i$ represent the corresponding $x$-set and $y$-set in the $i$-th row) if and only if $A(y_i)^T = c$ and each member of $y_i$ falls into the corresponding range $d_i$ from $d$. As we have showed above, $\varphi_1,..., \varphi_n$ are independent in private case. By POSTULATE 1, each of these functions should be considered by the solver separately from the others.

This last remark completes the natural proof for P ≠ NP.

## 6. CONCLUSION

The proof of P≠NP given in the article is not purely mathematical, since it uses, in any case, one purely physical postulate about the limited capacity of the Turing machine.

An obvious connection with the postulate of the theory of relativity in connection with the limited speed of light can be found if we draw an analogy with the emission of a photon of light and the operation of transition in a Turing machine (in this case, the speed of light is a physical analogue of the speed of information processing by a Turing machine).

The postulates we have introduced characterize the understanding of complexity that is intuitively used by algorithm developers. Therefore, we are not talking about a "universal" mathematical proof of the P≠NP formula, but about a proof within the framework of the accepted postulates.